\pdfoutput=1

\documentclass[11pt]{article}

\usepackage[preprint]{acl}

\usepackage{times}
\usepackage{latexsym}

\usepackage[T1]{fontenc}

\usepackage[utf8]{inputenc}

\usepackage{microtype}

\usepackage{inconsolata}

\usepackage{graphicx}

\usepackage{algorithm}
\usepackage{algpseudocode} 
\usepackage{caption} 
\usepackage{amsfonts} 
\usepackage{booktabs}
\usepackage{multirow}
\usepackage{enumitem}
\usepackage{amsmath}

\newcounter{todocounter}
\usepackage{etoolbox}
\usepackage{environ}

\title{SimAug: Enhancing Recommendation  with Pretrained Language Models for Dense and Balanced Data Augmentation}

\author{Yuying Zhao, Xiaodong Yang, Huiyuan Chen, Xiran Fan, Yu Wang, Yiwei Cai, Tyler Derr  \\
\{yuying.zhao, yu.wang.1, tyler.derr\}@vanderbilt.edu; \{xiaodyang, hchen, xirafan, yicai\}@visa.com}

\begin{document}
\maketitle
\begin{abstract}

Deep Neural Networks (DNNs) are extensively used in collaborative filtering due to their impressive effectiveness. These systems depend on interaction data to learn user and item embeddings that are crucial for recommendations. However, the data often suffers from sparsity and imbalance issues: limited observations of user-item interactions can result in sub-optimal performance, and a predominance of interactions with popular items may introduce recommendation bias. To address these challenges, we employ Pretrained Language Models (PLMs) to enhance the interaction data with textual information, leading to a denser and more balanced dataset. Specifically, we propose a simple yet effective data augmentation method (\textit{SimAug}) based on the textual similarity from PLMs, which can be seamlessly integrated to any systems as a lightweight, plug-and-play component in the pre-processing stage. Our experiments across nine datasets consistently demonstrate improvements in both utility and fairness when training with the augmented data generated by \textit{SimAug}. The code is available at \url{https://github.com/YuyingZhao/SimAug}.

\end{abstract}

\section{Introduction}
Pretrained Language Models (PLMs), trained on vast datasets, encapsulate a deep understanding of open-world knowledge, exhibiting substantial reasoning and generalization capabilities~\cite{brown2020language}. Their application across various fields has led to significant successes~\cite{hadi2023survey}. In this work, we investigate the potential of PLMs to enhance recommendation systems, especially focusing on improving the data quality. Recommender systems aim to provide personalized recommendations of items that a user might be interested in on online platforms, thereby mitigating information overload issues~\cite{ko2022survey}. These systems typically learn user/item embeddings from historical interactions. Traditionally, recommender systems have predominantly utilized ID-based embeddings, where users and items are represented by unique identifiers and these identifiers are then transformed into embedding vectors as learnable parameters~\cite{yuan2023go}. This approach has been well-established and widely-used in practice.

Despite the success of ID-based recommendation models, their heavy reliance on the interaction inherently limits their effectiveness by the quality of the interaction data.
When the interaction data is sparse, there is insufficient training data to update the systems, resulting in sub-optimal performance~\cite{al2018reducing}. Additionally, the real-world interaction data is often imbalanced with a predominance of interactions with popular items~\cite{abdollahpouri2019managing}. This will result in biased recommendations when ID-based models are trained on such skewed datasets. 
To address these challenges, researchers have incorporated auxiliary data sources like attributes~\cite{chen2020esam} and tags~\cite{zhang2024collaborative} into the models. Since this data is often in the format of text and recent breakthroughs in PLMs offer exciting prospects for enhancing recommendation systems~\cite{lin2023can}, new opportunities have emerged to enhance recommendation systems. 

Recent studies have taken innovative approaches to use PLMs to overcome data sparsity and bias.
For example, \cite{wang2024large} utilize prompts to generate synthetic data that encourages the exploration of less popular items, which is subsequently integrated into the training process through auxiliary pairwise loss functions. \cite{huang2024large} combines textual embeddings from PLMs with interaction embeddings from traditional recommendation and perform a further refinement by prompting the PLMs. 
These approaches underscore the potential of PLMs in enhancing recommendation. However, they predominantly rely on the prompt-based paradigm, which can be inefficient for processing long sequences. Additionally, it may require heavy manual trail-and-error to design the best prompt and the performance could be highly sensitive to the PLM choices.
Moreover, the integration of augmented data with complex model designs, such as new loss functions, often complicates the assessment of whether improvements arise from enhanced data quality or other modifications. In this study, we aim to explore alternative lightweight methods for extracting knowledge from PLMs, focusing exclusively on dataset enhancements to isolate and examine the impact of data quality on model performance.

\begin{figure}[t]
    \centering
\includegraphics[width=1\linewidth]{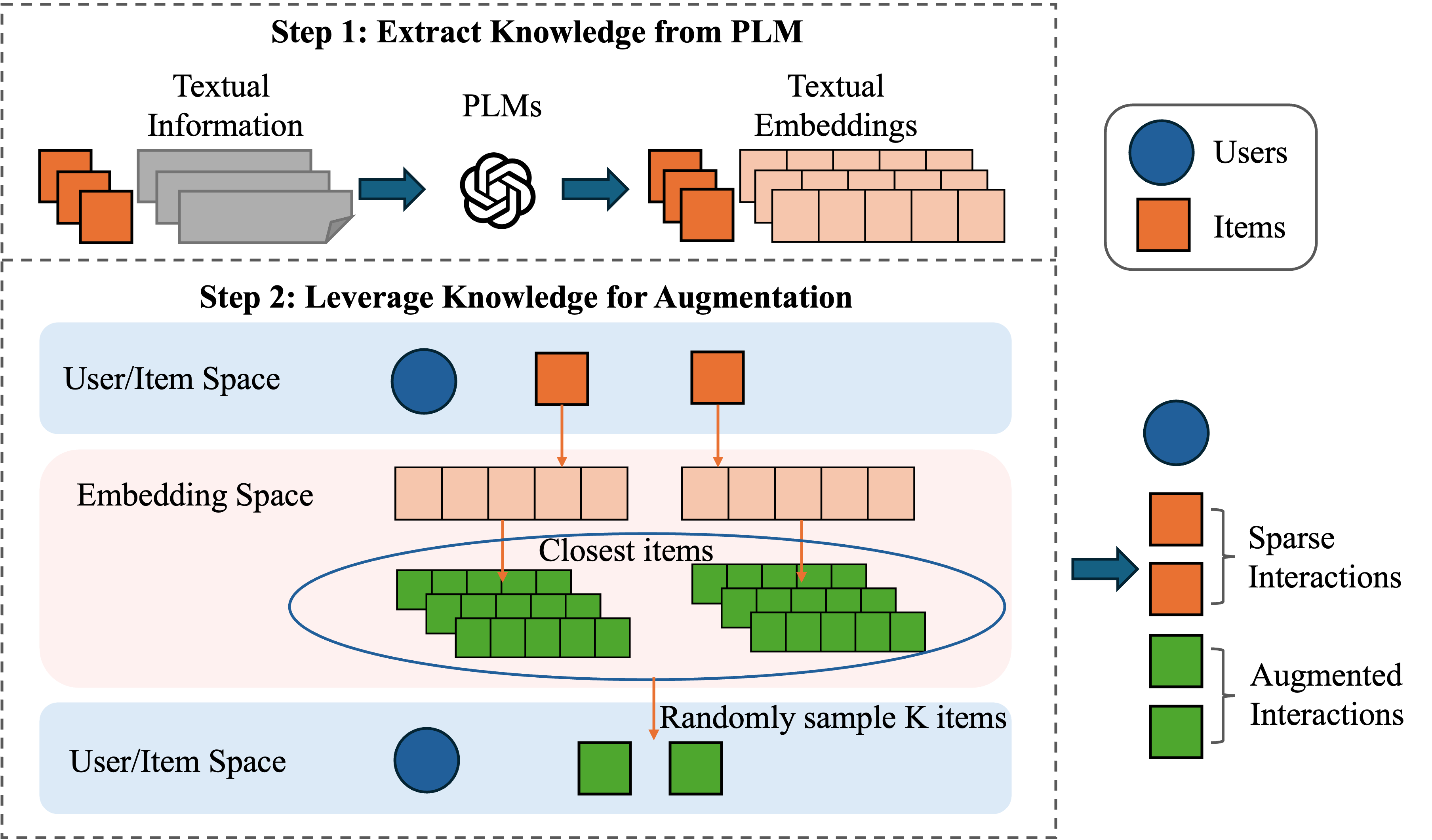}
    \caption{Two-step \textit{SimAug} framework for interaction augmentation based on Pretrained Language Models (PLMs). The knowledge from PLM is first extracted in the textual embeddings and then interactions are augmented based on the embeddings.}
    \label{fig.method}
\end{figure}

In this paper, we first empirically investigate the sparsity and imbalance issues in real-world datasets and dive into their relationship with the model performance in terms of the utility and fairness aspects. Thereafter, to mitigate the data sparsity and imbalance issues, we employ Pretrained Language Models (PLMs) to enhance the interaction data with textual information, obtaining a denser and more balanced dataset. Specifically, we propose a simple yet effective data augmentation method (\textit{SimAug}) based on the similarity in the textual embedding space obtained from PLMs (Fig.~\ref{fig.method}). The proposed \textit{SimAug} is lightweight and only needs to obtain the textual embedding once without repeatedly conducting inference for different users. We conduct extensive experiments to empirically investigate the impact of the augmented dataset and compare \textit{SimAug} with other augmentation methods (e.g, based on features, and pre-trained recommender systems) and other variants (e.g., based on similar users and different PLMs). Although empirical evidence suggests that PLMs may contain biases due to training on biased data~\cite{wang-etal-2024-large,10.1162/coli_a_00524}, our work demonstrates that PLMs can help mitigate bias in recommendations.

In summary, this work makes the following key contributions:
\begin{itemize}[leftmargin=*]
    \item We investigate and mitigate two key data issues in recommendation: sparsity and imbalance.
    \item We propose a simple yet effective augmentation framework \textit{SimAug} that contributes to a denser and more balanced dataset.
    \item Extensive experiments over various datasets verify the effectiveness of \textit{SimAug} in terms of improving utility and fairness performance.
\end{itemize}

\section{Data Sparsity and Imbalance}
\label{sec.issues}

In this section, we empirically investigate the relationship between sparsity/density and utility performance, and the relationship between data imbalance and fairness performance. We conduct experiments over 9 datasets (denoted from D1 to D9 in this section) and the detailed dataset descriptions are provided in Sec.~\ref{sec.dataset}.

\begin{figure}[h]
    \centering
    \includegraphics[width=1\linewidth]{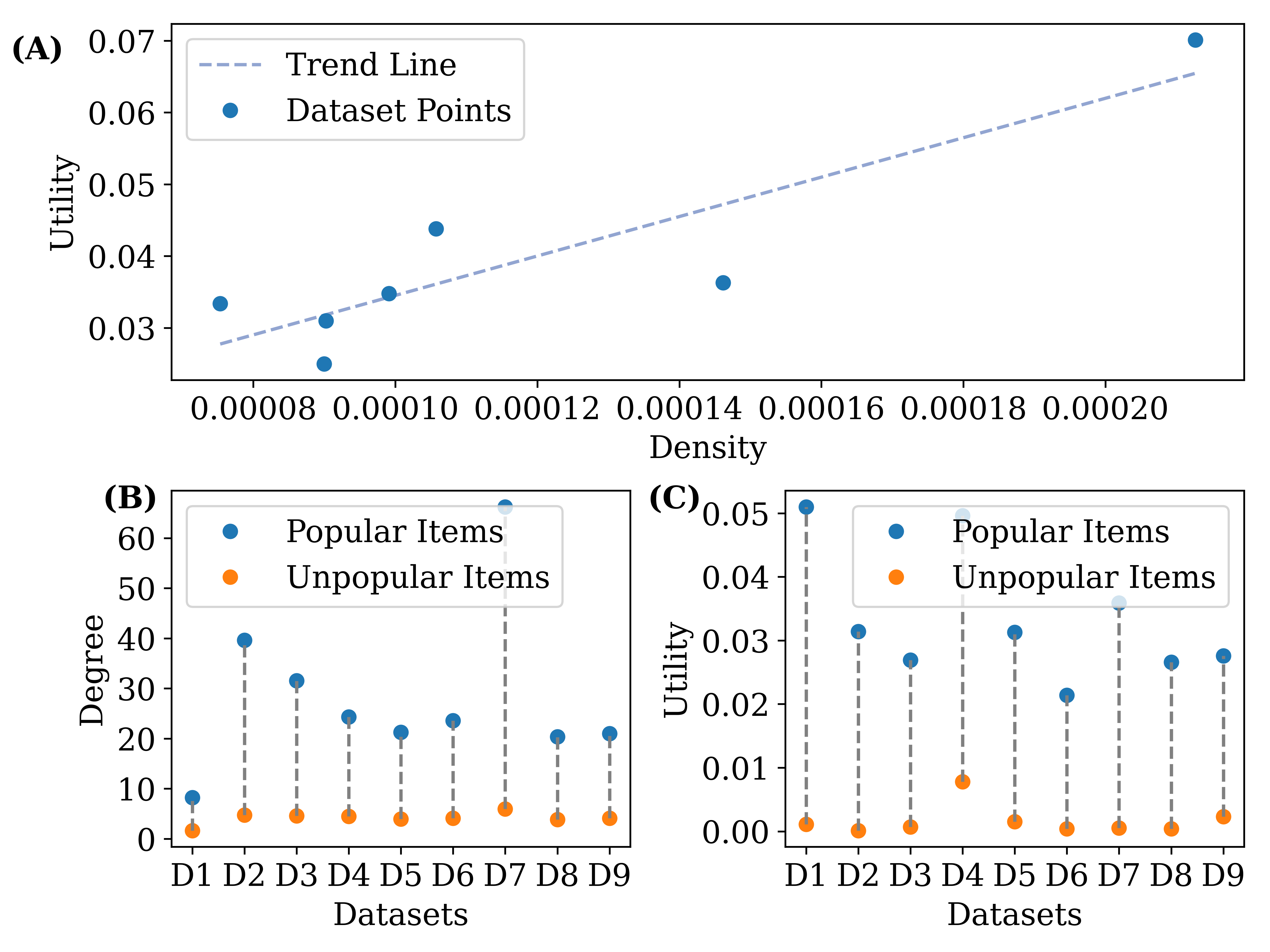}
    \caption{Investigating data sparsity and imbalance issues in real-world datasets: (A) utility and density, (B) average degree of popular and unpopular items, (C) utility performance of popular and unpopular items. Vertical dashed line in (B)(C) means that popular items have a higher score than the unpopular items.}
    \label{fig.prelim}
\end{figure}

\subsection{Data Sparsity}
Existing research has highlighted the challenge of sparse data in collaborative filtering where sparse interactions can seriously affect the recommendation quality of recommender systems~\cite{chen2018survey}. To investigate sparsity, we focus on the large-scale datasets with over 1 million edges. The density is calculated as: $\text{Density} = \frac{\# \text{Interactions}}{\# \text{Users} \#\text{Items}}$. We obtain the recommendation performance trained based on the classical model LightGCN~\cite{he2020lightgcn} and plot the results in Fig.~\ref{fig.prelim}(A). As shown in the figure, under the same  backbone, the recommendation performance vary across different datasets. The model trained on the dataset with a higher density tends to have a better utility performance. The correlations between them according to different metrics are as follows (the first value is the correlation and the second is p-value): (1) \textit{Pearsonr correlation}: (0.8871, 0.0077), (2) \textit{Spearmanr correlation}: (0.8571, 0.0137), (3) \textit{Kendalltau correlation}: (0.7143, 0.0302). The results indicate that there is a strong positive correlation between model performance and data density.
A denser dataset typically facilitates higher recommendation performance. Conversely, a sparse dataset is more challenge to learn. This can hinder the model's effectiveness, restricting its ability to generate accurate recommendations.

\subsection{Data Imbalance}
In classification scenarios, datasets often exhibit label imbalance, such as a predominance of positive labels over negative ones~\cite{thabtah2020data,yang2024simce,chen2021structured}. This issue is also prevalent in recommendation datasets, where interactions between users and items are crucial. 
User-item interactions are significantly higher between users and popular items compared to those with unpopular ones, leading to dataset imbalance.
We follow \cite{abdollahpouri2019managing} in popular/unpopular item division and find that popular items often dominate these interactions (Figure~\ref{fig.prelim}(B)). Their higher degree signifies greater user engagement with the popular item. The model trained on the biased data will have biased recommendation performance~\cite{zhao2024can,lai2023enhancing,wang2022improving}. As presented in Figure~\ref{fig.prelim}(C), popular items consistently outperform unpopular ones in the utility performance, indicating that the popular items will be recommended more accurately. Consequently, even if a user has an interest in less popular items, the recommendation model may fail to accurately predict and recommend these items. Over time, this imbalance will be enlarged, negatively affecting economic development~\cite{ahanger2022popularity}.
\section{SimAug: Interaction Augmentation}
\label{sec.method}
The issues of data sparisty and imbalance hinder the development of high-quality and fair recommendation systems. To mitigate these issues, we propose the \textit{SimAug} solution to augment the user-item interactions. In this section, we first introduce the motivation of using Pretrained Language Models for augmentation and then dive into the specifics of the proposed method.

\subsection{Motivation}
Real-world  datasets often face the data sparsity and imbalance issues. To mitigate these issues, researchers have looked into how to augment the incomplete interactions. However, since the original interaction distribution is imbalanced, directly relying on the biased data for generation might introduce extra bias. Therefore, researchers have investigated auxiliary information such as attributes~\cite{chen2020esam} and tags~\cite{zhang2024collaborative}. Since these features are mainly in the format of text, it is natural to use a pretrained language model to process them considering the significant success of PLMs in various applications~\cite{hadi2023survey}. Furthermore, since PLMs embody a comprehensive understanding of the world, they might introduce additional benefits (e.g., the understanding of the relationships among items).
\textit{Can we use pretrained language models to overcome the data quality issue in recommendation domain?} Answering this question will not only benefit the recommendation task, but also provide insights on other tasks.

\subsection{Method}
We propose a \underline{Sim}ple and straightforward similarity-based interaction \underline{Aug}mentation framework (\textit{SimAug}), which is illustrated in Fig.~\ref{fig.method} and its pseudocode is presented in Algorithm~\ref{alg.method}. It follows a two-step process. \textit{Firstly, we extract the knowledge from pretrained language models.} Textual information associated with items (e.g., titles, metadata, reviews) is processed by language models to generate textual item embeddings. \textit{Secondly, we leverage the extracted knowledge for augmentation.} 
Since inactive users and unpopular items are most affected by data sparsity issue, we focus on augmenting interactions between them. For active and inactive user partition, we follow \cite{li2021user}. Specifically, for the inactive users, we aim to augment their interactions based on their limited historical interactions. For each of the interacted items, we identify the top $k$ similar items from the unpopular item set in the textual embedding space and randomly select $K$ items to form the augmented interactions (from line 10-17). 

Notably, textual embeddings are only computed once and can be cached for future use, enhancing the efficiency. Furthermore, our \textit{SimAug} only modifies the original data without imposing requirements on downstream models, making it easy to be implemented with existing backbones.

\begin{algorithm}[H]
\small
\caption{\textit{SimAug} Framework}
\label{alg.method}
\begin{algorithmic}[1]
\State \textbf{Input:} Pretrained language model $\mathcal{L}$, textual data of items \textit{text}, historical interactions $\mathcal{H}$, inactive users $\mathcal{U}_\text{inactive}$, unpopular items $\mathcal{I}_{\text{unpopular}}$, similarity threshold $k$, number of augmented items $K$
\State \textbf{Output:} Augmented interaction

\State \underline{Step 1: Knowledge Extraction}
\For{each item $i$}
    \State $\mathbf{e}_i \gets \mathcal{L}(\text{text}[\text{i}])$ 
\EndFor
\State Cache all item embeddings $\mathcal{E} = \{\mathbf{e}_i\}, $ 

\State \underline{Step 2: Augmentation}

\State $\mathcal{D}_{\text{aug}} = \{\}$ 

\For{user $u \in \mathcal{U}_\text{inactive}$}
    \State $\mathcal{C}_u = \{\}$
    \For{item $i \in \mathcal{H}_u$}
        \State $\mathcal{C}_i^k = \{ j_1, \dots, j_k \mid j \in \mathcal{I}_{\text{unpopular}},\newline \hspace*{14ex} \text{rank}(\text{sim}(\mathbf{e}_i, \mathbf{e}_j)) \leq k \}$
        \State $\mathcal{C}_u = \mathcal{C}_u \bigcup \mathcal{C}_i^k$
    \EndFor
    \State $\mathcal{D}_\text{aug} = \mathcal{D}_\text{aug} \cup \{(u, a) \mid a \in \text{RandomSample}(\mathcal{C}_u,k)\}$
    
\EndFor
\State \textbf{Return} $\mathcal{D}_\text{aug}$

\end{algorithmic}
\end{algorithm}
\section{Experiments}
\label{sec.exp}

In this section, we evaluate the performance of \textit{SimAug} on real-world datasets. We aim to answer the following research questions:
\begin{itemize}[leftmargin=*]
    \item \textbf{RQ1:} Does \textit{SimAug} boost the performance of vanilla models and how effective it is when compared with other augmentation methods?
    \item \textbf{RQ2:} Which way is more effective when augmenting interactions based on users or items?
    \item \textbf{RQ3:} Whether interaction augmentation is more effective than feature augmentation?
    \item \textbf{RQ4:} How does the choice of PLMs affect the performance?
\end{itemize}

\subsection{Experimental Settings}
\subsubsection{Datasets}
\label{sec.dataset}
We conduct extensive experiments on 9 datasets from the publicly available amazon repository\footnote{Amazon Review'23: \newline \hspace*{3ex} https://amazon-reviews-2023.github.io/main.html}~\cite{hou2024bridging}. This repository contains rich interaction and textual information. If not specified, we only use the title of the items as the text input to the language models while other information such as the reviews or descriptions can also be potentially used. We perform the following pre-processing procedure: items without title are filtered out, additionally we perform 5-core filtering which is commonly used in recommendation systems. The dataset statistics are in Appendix~\ref{app.exp}.

\subsubsection{Recommender Backbones}
In this paper, we conducted comprehensive experiments over LightGCN~\cite{he2020lightgcn}, due to its superior performance. 

\subsubsection{Pretrained Language Models}
We use the sentence transformer model \textit{all-MiniLM-L6-v2} as the PLM. Other PLMs including miniLM and LLMs are explored in Sec.~\ref{sec.lms}.

\subsection{Evaluation Measurements}

We assess the models trained on the datasets, both pre- and post-augmentation, examining their utility and fairness performance. For utility, we measure five metrics: Recall@20, NDCG@20, F1@20, Precision@20, and HitRate@20. Due to space constraints, we report only Recall@20, NDCG@20, and Avg@20 (i.e., the average of all five metrics). In terms of fairness, we analyze the average utility metrics for both popular and unpopular items (denoted as Pop@20 and Unpop@20), defining popularity based on the number of interactions as per \cite{abdollahpouri2019managing}. We also calculate item fairness, quantified as the ratio of utility performance for unpopular items to that for popular items. Since the popular items tend to outperform the unpopular ones even after the mitigation, the item fairness score falls in the range between 0 and 1. It is the higher the better. We repeat the experiments three times and report the average. We have omitted the standard deviation from the table, as it is often negligible.

\subsubsection{Compared Methods}
We compare \textit{SimAug} with the vanilla model trained on the original dataset and other augmentation strategies. \textit{Aug-Random} randomly augments the interactions by selecting items from all items in the training data for each inactive users. To enable knowledge extraction from pre-trained models, we compare the variants based on pre-trained recommender to compare the benefits of integrating different source of information (i.e., text and recommendation).
\textit{Aug-Rec}  selects the augmented items based on the pre-trained item embeddings from recommendation. Similar to \textit{SimAug}, the top closest items in the recommendation embedding space are firstly obtained to form the candidate pool and then $K$ items are randomly selected from the pool for augmentation. Different strategies add the same number of interactions for a fair comparison. Additionally, we compare (1) \textit{SimAug} variants to augment interactions based on users in Sec.~\ref{sec.user_item}, (2) other ways of using textual information in Sec.~\ref{sec.feature_aug}, and (3) different variants under \textit{SimAug} framework by changing the language models in Sec.~\ref{sec.lms}. 

\subsection{RQ1: Effectiveness of \textit{SimAug}}

\subsubsection{Utility Performance}

\setlength{\tabcolsep}{2.1pt}
\begin{table*}[t]

\caption{Utility performance with LightGCN backbone. The improvement compares \textit{SimAug} to the vanilla.}
\label{tab.utility_results}
\resizebox{\textwidth}{!}{%
\begin{tabular}{l|ccc|ccc|ccc}
\toprule
\multirow{2}{*}{\textbf{Method}} &          \textbf{Recall@20}   & \textbf{NDCG@20}   &\textbf{Avg@20}   & \textbf{Recall@20}   & \textbf{NDCG@20}   &\textbf{Avg@20} & \textbf{Recall@20}   & \textbf{NDCG@20}   &\textbf{Avg@20}  \\ 
\cline{2-10}
         & \multicolumn{3}{c|}{Appliances} & \multicolumn{3}{c|}{Baby Products} & \multicolumn{3}{c}{Grocery and Gourmet Food} \\ 
       
\hline\hline  Vanilla  &  0.0692 & 0.0328 & 0.0369  & 0.0477 & 0.0230 & 0.0365 & 0.0434 & 0.0234 & 0.0348\\ 
Aug-Random     & 0.0482 & 0.0194 & 0.0249 & 0.0459 & 0.0226 & 0.0354 & 0.0357 & 0.0192 & 0.0292\\ 
Aug-Rec     & 0.0635 & 0.0338 & 0.0344 & 0.0459 & 0.0248 & 0.0354 & 0.0424 & 0.0237 & 0.0342\\ 
SimAug      &  0.0983 & 0.0424 & 0.0515 & 0.0546 & 0.0283 & 0.0419 & 0.0514 & 0.0286 & 0.0408\\ 
Improve (\%) & \textbf{+42.07\%}  & \textbf{+29.09\%}  & \textbf{+39.71\%} & \textbf{+14.55\%} &  \textbf{+22.85\%} &  \textbf{+14.78\%} &\textbf{+18.34\%} &\textbf{+22.29\%}& \textbf{+17.09\%} \\ \hline\hline

&          \multicolumn{3}{c|}{Movies and TV} & \multicolumn{3}{c|}{Office Products} & \multicolumn{3}{c}{Patio Lawn and Garden} \\
\hline   Vanilla & 0.0898 & 0.0554 & 0.0701 & 0.0485 & 0.0268 & 0.0363 & 0.0328 & 0.0172 & 0.0250\\ 
Aug-Random     &  0.0943 & 0.0607 & 0.0754 & 0.0387 & 0.0212 & 0.0294 & 0.0258 & 0.0135 & 0.0201\\ 
Aug-Rec & 0.1008 & 0.0556 & 0.0779 & 0.0487 & 0.0268 & 0.0365 & 0.0319 & 0.0169 & 0.0244\\
SimAug  &  0.1149 & 0.0675 & 0.0886 & 0.0535 & 0.0292 & 0.0399& 0.0386 & 0.0208 & 0.0294\\ 
Improve (\%) & \textbf{+27.95\%} & \textbf{+21.86\%} & \textbf{+26.40\%} & \textbf{+10.21\%} & \textbf{+8.84\%}  & \textbf{+9.94\%}  & \textbf{+17.82\%} & \textbf{+20.44\%} & \textbf{+17.54\%}\\ 

\hline \hline 

&          \multicolumn{3}{c|}{Pet Supplies} & \multicolumn{3}{c|}{Sports and Outdoors} & \multicolumn{3}{c}{Toys and Games} \\
\hline   Vanilla & 0.0559 & 0.0291 & 0.0438 & 0.0403& 0.0213 &0.0310 & 0.0438 & 0.0244 & 0.0334\\ 
Aug-Random     &   0.0490 & 0.0256 & 0.0388 & 0.0319 & 0.0169 & 0.0250 & 0.0326 & 0.0185 & 0.0255\\ 
Aug-Rec & 0.0552 & 0.0306 & 0.0435& 0.0367 & 0.0203 & 0.0284 & 0.0362 &0.0200  & 0.0280\\
SimAug  &   0.0589 & 0.0313 & 0.0461 & 0.0428 & 0.0228 & 0.033 & 0.0465 & 0.0253 & 0.0352\\ 
Improve (\%) &  \textbf{+5.37\%} & \textbf{+7.52\%} & \textbf{+5.32\%} & \textbf{+6.00\%} &  \textbf{+7.32\%} & \textbf{+6.19\%} & \textbf{+6.12\%} & \textbf{+3.67\%} & \textbf{+5.43\%}\\ \toprule
\end{tabular} 
}
\end{table*} 

Table~\ref{tab.utility_results} shows the utility performance across 9 datasets for the LightGCN backbones. The improvement over vanilla model (i.e., no augmentation) is calculated and highlighted in bold. From the result, we draw the following observations.
\begin{itemize}[leftmargin=*]
    \item Simply addressing data sparsity does not guarantee improved performance, as demonstrated by the random strategy. This approach increases data density by adding more interactions but results in performance degradation when compared to models trained on the pre-augmentation dataset. Randomly inserting interactions, if not carefully considered, introduces noise, resulting in a deterioration in performance.
    \item Augmentation based on the pre-trained recommendation embeddings shows variable stability across different datasets. For instance, the performance remains steady for \textit{Office Products}, declines for \textit{Appliances}, and improves for \textit{Movies and TV}. Inactive users, with only limited historical interactions available, present challenges in accurately inferring their interests for recommendations, leading to sub-optimal performance. Consequently, relying on recommendation embeddings for data augmentation tends to yield results that are sub-optimal and unstable.
    \item Our method achieves a consistent performance gain over the vanilla one. The improvement in the average score ranges from $5\%$ to around $40\%$. This validates the effectiveness of the pretrained language model based augmentation.
\end{itemize}

\subsubsection{Fairness Performance}
Table~\ref{tab.fairness_results} shows the fairness performance. Since the random-based strategy already shows a poor utility performance in the previous discussion, we exclude it for the further fairness evaluation. We draw the following observations:
\begin{itemize}[leftmargin=*]
    \item The recommendation based on the original dataset suffer from severe unfairness. The recommendation for popular items are much better than unpopular ones. In some dataset (e.g., Baby Products), the recommendation is unable to work for unpopular items due to the imbalanced data.
    \item The augmentation based on recommendation embedding space will exaggerate the unfairness, evidenced by the smaller fairness scores across the datasets. Since the recommendation model trained on imbalanced data tend to favor towards popular items, it results in a worse performance for recommending unpopular items accurately. This indicates that the augmentation based on biased knowledge would enlarge the bias.
    \item Our augmentation method contributes to a more balanced dataset, thereby enhancing fairness. From the results, it is evident that \textit{SimAug} generally improves utility for both popular and unpopular items. Since unpopular items typically suffer more from poor performance due to limited interactions, they experience a more significant improvement from the inclusion of textual content. This substantial benefit leads to an overall improvement in the fairness.

\end{itemize}

\setlength{\tabcolsep}{2.1pt}
\begin{table*}[t]

\caption{Fairness performance with LightGCN backbone. The improvement compares \textit{SimAug} to the vanilla.}
\label{tab.fairness_results}
\resizebox{\textwidth}{!}{%
\begin{tabular}{l|ccc|ccc|ccc}
\toprule
\multirow{2}{*}{\textbf{Method}} &          \textbf{Pop@20}   & \textbf{Unpop@20}   &\textbf{Fairness}   & \textbf{Pop@20}   & \textbf{Unpop@20}   &\textbf{Fairness} & \textbf{Pop@20}   & \textbf{Unpop@20}   &\textbf{Fairness} \\ 
\cline{2-10}
         & \multicolumn{3}{c|}{Appliances} & \multicolumn{3}{c|}{Baby Products} & \multicolumn{3}{c}{Grocery and Gourmet Food} \\ 
       
\hline\hline  Vanilla  &  0.0510 & 0.0011 & 2.0638 & 0.0314 & 0.0001 & 0.1794  & 0.0269 & 0.0007 & 2.5431\\ 
Aug-Rec     & 0.0479 &0.0001 &0.2211& 0.0301 &0.0000 &0.013 &0.0263& 0.0004 &1.3366 \\ 
SimAug      & 0.0626& 0.0215 &34.3659 & 0.0357 &0.0021 &5.742 &0.0306 &0.0031& 10.2771\\ 
Improve (\%) & \textbf{+22.69\%} & \textbf{+1942.96\%} &\textbf{+1565.18\%} &\textbf{+13.96\%}&\textbf{+3547.44\%} &\textbf{+3100.60\%} &\textbf{+14.03\%} &\textbf{+360.79\%} &\textbf{+304.11\%}\\ \hline\hline

&          \multicolumn{3}{c|}{Movies and TV} & \multicolumn{3}{c|}{Office Products} & \multicolumn{3}{c}{Patio Lawn and Garden} \\
\hline   Vanilla & 0.0496 & 0.0078& 15.7805 &0.0313& 0.0015 &4.6585 &0.0214 &0.0004 &2.0732\\ 
Aug-Rec & 0.0549 &0.0061 &11.1712 &0.0314& 0.0010& 3.0524 &0.0208& 0.0001 &0.6651\\
SimAug  & 0.0600& 0.0190 &31.6080 &0.0332 &0.0049 &14.8501& 0.0248& 0.0015 &6.115\\ 
Improve (\%) & \textbf{+20.96\%} &\textbf{+142.28\%} &\textbf{+100.30\%} &
 \textbf{+6.15\%} &\textbf{+238.39\%}& \textbf{+218.77\%}& \textbf{+16.09\%}& \textbf{+242.41\%} &\textbf{+194.96\%} \\ 

\hline \hline 

&          \multicolumn{3}{c|}{Pet Supplies} & \multicolumn{3}{c|}{Sports and Outdoors} & \multicolumn{3}{c}{Toys and Games} \\
\hline   Vanilla & 0.0359 & 0.0005& 1.4791 &0.0266 &0.0004 &1.6130& 0.0276 &0.0023 &8.2593\\ 
Aug-Rec &0.0353 &0.0004& 1.0966 &0.0242 &0.0001& 0.5937& 0.0229 &0.0005 &2.2025\\
SimAug  &  0.0378 &0.0009 &2.4665 &0.0278 &0.0015 &5.4730 &0.0267 &0.0088 &32.9507\\ 
Improve (\%) &  \textbf{+5.26\%} &\textbf{+75.53\%}& \textbf{+66.76\%} &\textbf{+4.52\%} &\textbf{+254.64\%} &\textbf{+239.30\%}& -3.31\% &\textbf{+285.74\%}& \textbf{+298.95\%} \\ \toprule
\end{tabular} 
}
\end{table*} 

\subsection{RQ2: User vs Item-based Interaction Augmentation}
\label{sec.user_item}

\begin{figure}[h]
    \centering
    \includegraphics[width=0.9\linewidth]{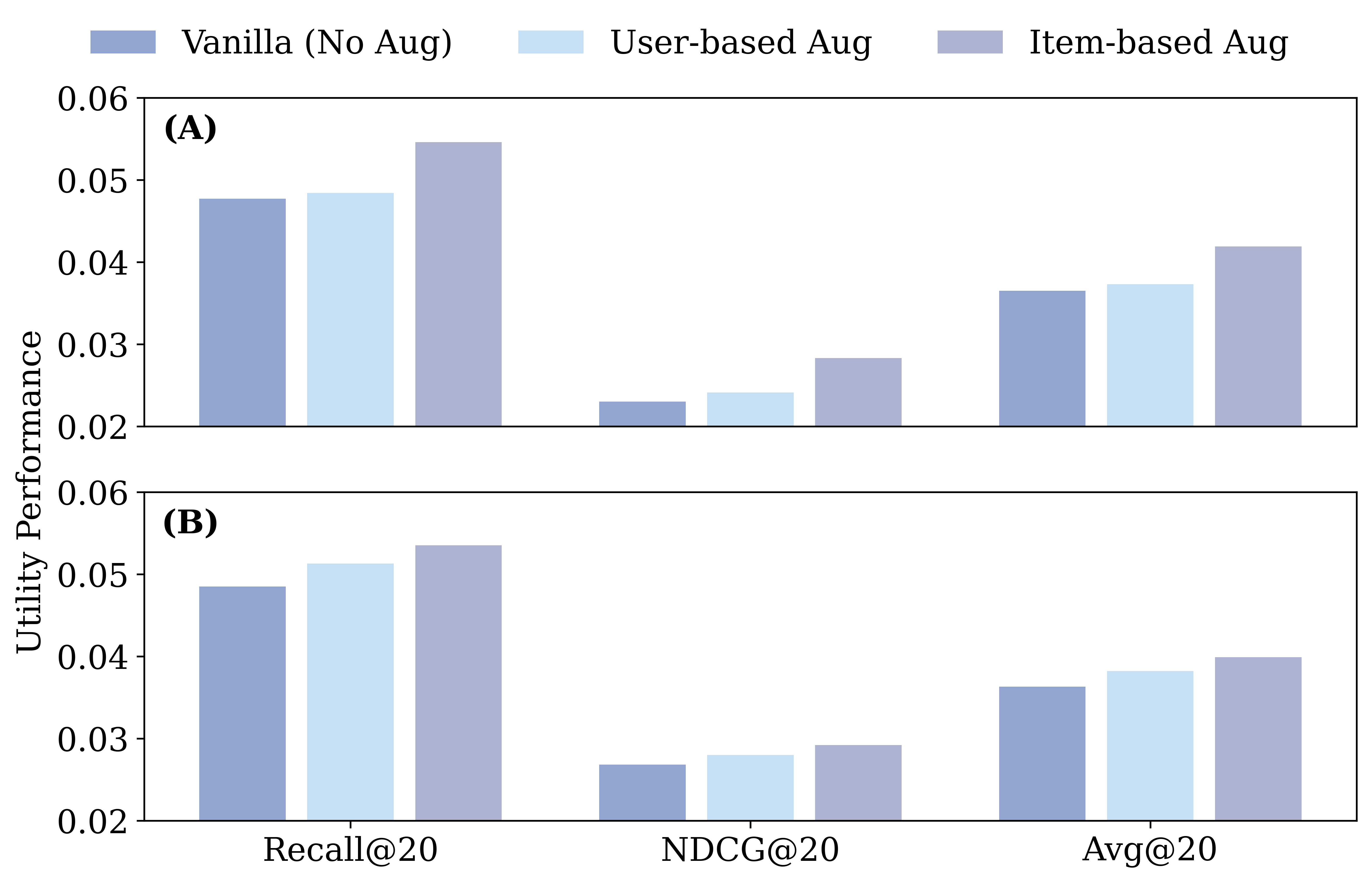}
    \caption{Comparison of user and item-based interaction augmentation (A) Baby Products, (B) Office Products.}
    \label{fig.user_item}
\end{figure}

\textit{SimAug} augments interactions based on the item similarity (Fig.~\ref{fig.method}). Due to the symmetric nature in recommendation, the interactions can also be augmented based on user similarity. Specifically, for unpopular items with sparse interactions, we can augment the interactions by adding close users with the historically interacted users. Since there is no inherent textual user features, we use the titles of interacted items to represent the textual information of users. Fig.~\ref{fig.user_item} shows the results on two datasets, indicating the following observations:
\begin{itemize}[leftmargin=*]
    \item Both augmentation strategies enhance performance compared to the pre-augmentation dataset, demonstrating the effectiveness of using a language model to incorporate textual knowledge.
    \item The item-based interaction augmentation is more effective than the user-based interaction augmentation in terms of improving the model performance. There are several potential reasons behind this: (1) The textual description of items are more informative than the users since the user description is based on the title of interacted items and lacks direct textual features; (2) Users, often having diverse interests, are more complex than items. Similar users identified in the embedding space may share similarities due to one interest but may not necessarily have consistent preferences for the same items; (3) To ensure the number of augmented edges to be at the same level, more edges are added per item for the user-based interaction due to smaller number of items than users. This might introduce more noise.  In summary, the user-based interaction augmentation might be less reliable than the item-based one. This necessitates additional design considerations for user-based interaction augmentation to achieve better performance. Similar distinctions between user and item complexities have been noted in \cite{mao2021ultragcn} for pure interactions.
\end{itemize}

\subsection{RQ3: Interaction vs Feature Augmentation}
\label{sec.feature_aug}
\begin{figure}
    \centering
    \includegraphics[width=0.9\linewidth]{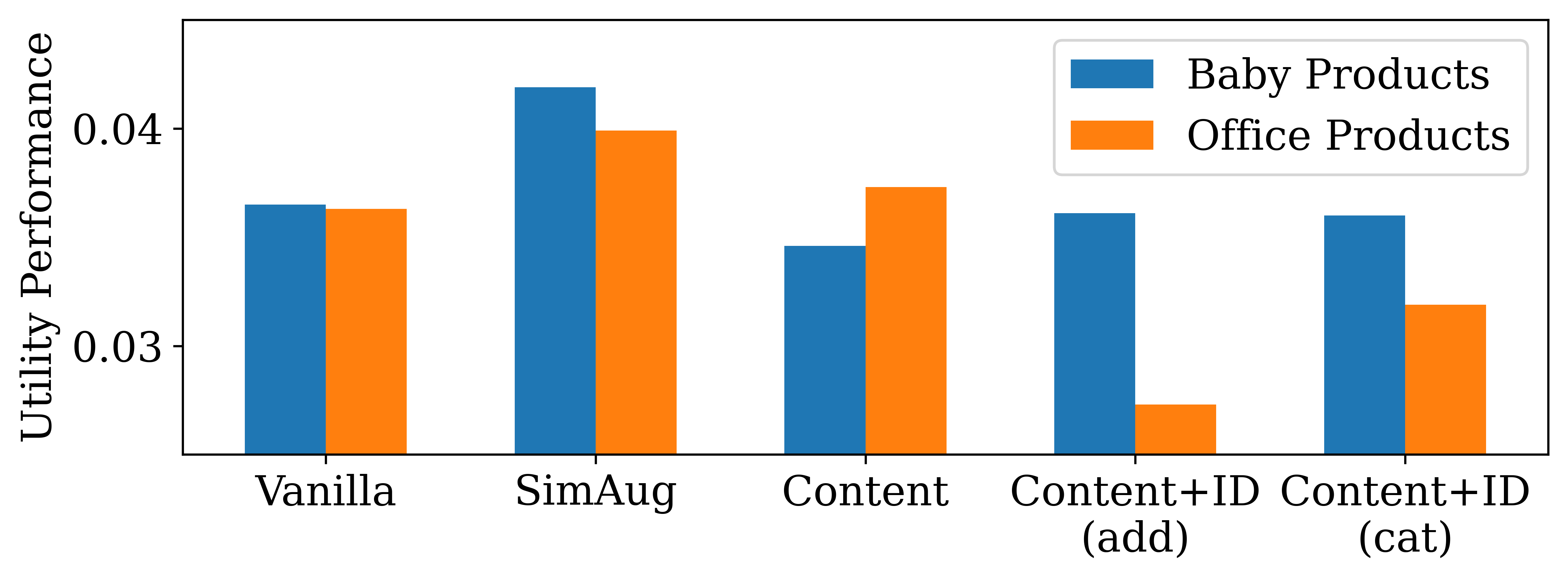}
    \caption{Comparison of other content-based variants.}    
    \label{fig.content_comparison}
\end{figure}

In addition to augmenting interactions, we compare the following ways to integrate the textual information~\cite{yuan2023go}:
\begin{itemize}[leftmargin=*]
    \item \textit{Content}: textual content are fed into PLMs to obtain the embeddings and then these embeddings are used as the user/item embeddings.
    \item \textit{Content+ID (add)}: user/item embeddings are a combination of the content embedding and learnable ID embeddings. Add denotes that the aggregation method here is addition.
    \item \textit{Content+ID (cat)}: similar to the above one, this one uses concatenation operation to aggregate the content and ID embeddings.
\end{itemize}

Results in Fig.~\ref{fig.content_comparison} show that simply integrating content into the embeddings cannot ensure the improvement of utility performance. The aggregation methods that use both textual embeddings and ID embeddings also do not show consistent improvement. Our proposed \textit{SimAug} is more effective in terms of using the textual information than these feature-based augmentation.

\begin{table*}[]
\centering
\small
\caption{Impact of different LLMs on Baby and Office Products.}
\label{tab:plms}
\begin{tabular}{l|c|ccc|ccc}
\toprule
&     \multirow{2}{*}{Model Size}  &    \textbf{Recall@20}   & \textbf{NDCG@20}   &\textbf{Avg@20}   & \textbf{Recall@20}   & \textbf{NDCG@20}   &\textbf{Avg@20} \\ 
\cline{3-8}
    &      & \multicolumn{3}{c|}{Baby Products} & \multicolumn{3}{c}{Office Products} \\ 
       
\hline 
Vanilla& - & 0.0477& 0.0230 &0.0365 &0.0485 &0.0268 &0.0363 \\
\hline
SBert-L6& 80MB & 0.0546 &0.0283& 0.0419 & 0.0535 & 0.0292& 0.0399\\
SBert-L12 &120MB &0.0538 &0.0277 &0.0413 & 0.0538& 0.0291 &0.0401\\
SBert-Mpnet &420MB &0.0534 &0.0274 &0.0410 & 0.0530 & 0.0289 &0.0396 \\
\hline
gte-small & 0.07GB & 0.0546 & 0.0280 &  0.0419 & 0.0553 &0.0304& 0.0411 \\
gte-base & 0.22GB & 0.0551 & 0.0284 & 0.0422 & 0.0554 & 0.0306 & 0.0413\\
gte-large & 0.67GB & 0.0546 & 0.0281 & 0.0419 & 0.0556 & 0.0306 & 0.0413 \\
\hline

Llama2-13b & 26GB &0.0519 &0.0265 &0.0399 & 0.0469 &0.0250 & 0.0352\\ 
\toprule

\end{tabular}
\end{table*}

\subsection{RQ4: Impact of PLMs}
\label{sec.lms}

In this section, we assess the impact of PLMs on the augmentation effectiveness by conducting experiments using various PLMs. For models specifically designed for embedding task (e.g., SBert and General Text Embeddings (GTE) model~\cite{li2023towards}), we directly obtained the textual embeddings. For large language models (LLMs) that are not tailored for embedding task, we followed \cite{jiang2023scaling} to extract their embeddings.

The results, shown in Table~\ref{tab:plms}, demonstrate a generally consistent performance improvement over the vanilla dataset, regardless of the specific PLM used for embedding extraction. This validates the robustness of the \textit{SimAug} framework across different PLMs. We also observe that model size has a minimal impact on performance gains. A larger model does not necessarily lead to greater improvement. In this case, the smaller model is sufficient for extracting textual embeddings for data augmentation, allowing for a more lightweight framework. LLMs, such as LLaMA 2~\cite{touvron2023llama}, do not exhibit significant improvements over smaller models. We hypothesize this is related to two factors: (1) the short length of the titles limits the utilization of the models' powerful capabilities, and (2) these models are not specifically trained for the embedding task, resulting in sub-optimal performance. The latter factor could inspire future research directions focused on extracting informative embeddings from LLMs.

\subsection{RQ5: Impact of Hyperparamter}
\label{sec.aug_num}
\begin{figure}
    \centering
    \includegraphics[width=0.9\linewidth]{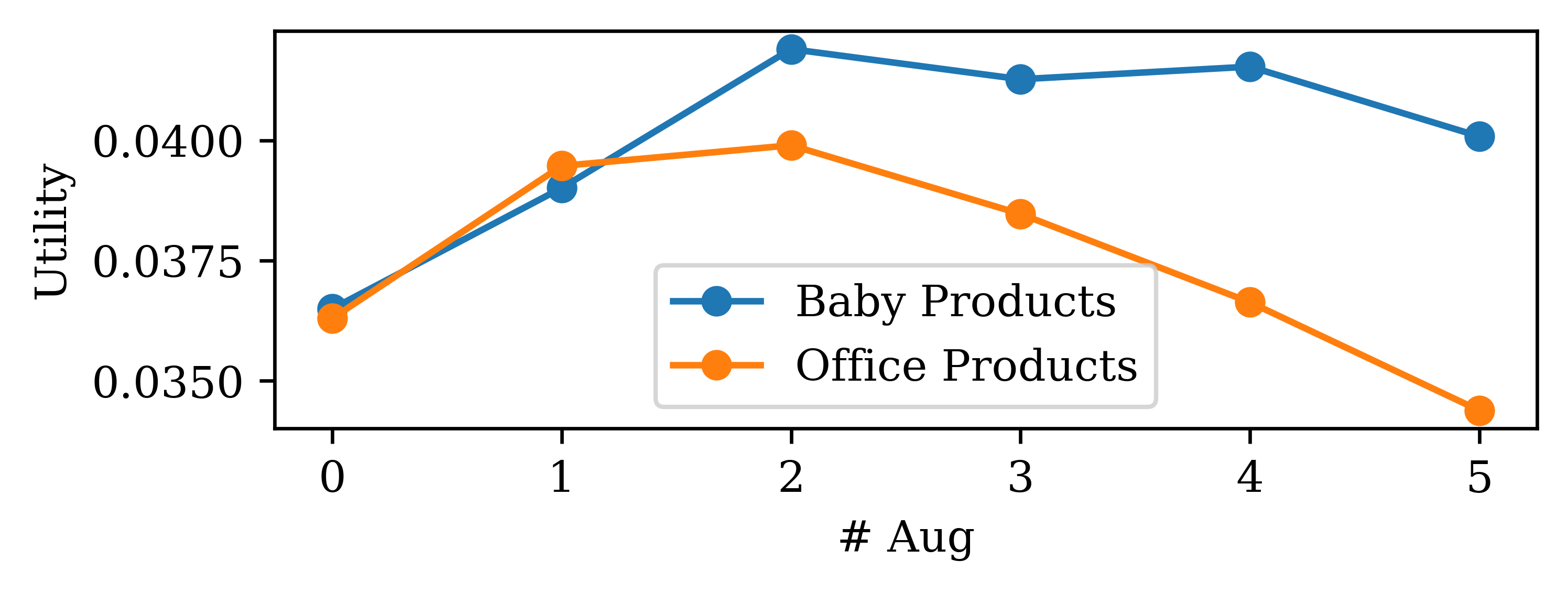}
    \caption{Impact of the number of augmentation.}
    \label{fig.aug_num}
\end{figure}

In this section, we discuss the effects of varying the number of augmentations per inactive user. As illustrated in Fig.~\ref{fig.aug_num}, the utility performance initially improves with an increase in the number of augmentations. This improvement underscores the advantages of utilizing augmentation techniques based on PLMs. However, beyond a certain threshold, increasing the number of augmentations does not necessarily yield better results. When excessively high, the number of augmentations can degrade the quality of top-ranked items, potentially introducing noise due to less relevant similarities.

\section{Related Works}
\label{sec.related}
\subsection{LM-enhanced Recommendation}
The potential of language models (LM) to enhance recommendation systems has been thoroughly explored in recent literature~\cite{wu2024survey,wang2025a}. A typical recommendation system encompasses six phases: data collection, feature engineering, feature encoding, scoring/ranking, user interaction, and recommendation pipeline management~\cite{sheng2025language, lin2023can}. LMs can be integrated into any of these phases. For example, LMs can serve as encoders to generate textual embeddings during the feature encoding stage~\cite{qiu2021u}, or they can be employed in the scoring stage to directly produce ratings or rankings instead of computing the scores based on the learned embeddings~\cite{tang2023one,bao2023tallrec}. This study primarily investigates LM-based data augmentation, focusing on addressing inherent dataset limitations in recommendation systems. 

\subsection{Fairness in Recommendation}

The imbalance issue in datasets, which can lead to unfair recommendations, is a significant concern. Research has revealed a bias wherein popular and unpopular items are treated disparately~\cite{klimashevskaia2024survey}. Various methods have been proposed to counteract this bias, including rebalancing techniques that generate resampling weights for selecting training samples~\cite{huang2006correcting}, and regularization-based methods that modify the optimization objective to penalize the recommendation of popular items~\cite{zhu2021popularity,chen2022denoising}. Additionally, auxiliary features such as attributes~\cite{chen2020esam} and tags~\cite{zhang2024collaborative} have been employed to mitigate bias resulting from interaction imbalances. 

The recent emergence of large language models (LLMs) has spurred research into leveraging them to enhance fairness in recommendation systems. Recent studies have explored innovative approaches: \cite{wang2024large} uses prompts to generate synthetic data from LLMs, asking users to choose between unpopular items, which is then incorporated into auxiliary pairwise loss functions. Meanwhile, \cite{huang2024large} combines textual embeddings from LLMs with interaction embeddings from traditional systems and perform a further refinement by prompting the LLMs. \cite{huang2024large} propose diverse prompts to assess the performance of LLMs as interaction simulators for cold-start item recommendations. \cite{wu2024coral} optimizes the retrieval stage to access minimally-sufficient collaborative information for recommendation tasks. These studies demonstrate the efficacy of utilizing LLMs for fair recommendation systems. However, prompt-based methods may face computational challenges and require labor-intensive design for specific prompts. Our research aims to explore alternative lightweight methods to extract knowledge from LLMs beyond the prompting paradigm. Additionally, we focus on the pre-processing stage, which is less explored in existing research, where in-processing methods significantly outnumber pre-processing ones.

\section{Conclusion}
In this work, we focus on enhancing the data quality in recommendation scenario by augmenting the sparse and imbalanced datasets to be a denser and more balanced one. To achieve this, we propose a PLM-based framework called \textit{SimAug}. This framework first extracts the knowledge from PLMs and augment the interactions based on the extracted knowledge in the textual embedding space. Extensive experiments validate the effectiveness of \textit{SimAug} in terms of improving utility and fairness performance. In the future, we plan to investigate the underlying reasons why the simple augmentation strategy enhances performance.

\newpage
\section*{Limitations}

Our study has conducted experiments on nine datasets that vary in density, but the impact on datasets beyond this density range has not been analyzed. Expanding the range of dataset densities could provide a deeper understanding of whether the augmentation framework is effective for extremely sparse or relatively dense datasets. Furthermore, we currently employ a single method for extracting textual embeddings from LLMs. This approach may not fully capture the range of possible embeddings, and results might differ with alternative extraction methods.  Our study focuses on the recommendation scenario and it would be intriguing to explore whether the framework can apply to other areas such as general textual graphs.

\bibliography{custom}

\appendix
\newpage
\section{Experiment Details}
\label{app.exp}

\subsection{Dataset Statistics}

The statistics of the datasets after pre-processing steps are shown in Table~\ref{tab.data}. 

\begin{table}[h]
\small
\caption{Dataset Statistics}
    \centering
    \resizebox{0.5\textwidth}{!}{%
    \begin{tabular}{l|c|c|c}
\toprule
        \textbf{Dataset} & \# \textbf{User} & \# \textbf{Item} & \# \textbf{Interaction} \\
        \hline
        Appliances & 18,830 & 6,542 &65,694\\
        Baby Products & 88,126  & 24,737 & 703,964\\
        Grocery and Gourmet Food &257,363 & 95,555 & 2,437,551\\
        Movies and TV & 106,074 & 54,679&1,233,612\\
        Office Products & 137,461 & 55,075&1,106,657\\
        Patio Lawn and Garden & 238,042 & 89,710& 1,921,639\\
        Pet Supplies & 365,763 & 80,194&3,101,564\\
        Sports and Outdoors & 215,038 & 91,719&1,780,007\\
        Toys and Games & 268,929 & 117,230 &2,375,168\\
      \toprule
    \end{tabular}
    }
    \label{tab.data}
\end{table}

The average degrees of the popular items (i.e., pop) and unpopular items (i.e., unpop) along with their ratio (i.e., Pop/Unpop) are presented in Table~\ref{tab:degree}.

\begin{table}[h]
    \centering
    \caption{Average degrees}
    \resizebox{0.5\textwidth}{!}{%
    \begin{tabular}{c|c|c|c}
    \toprule
        Dataset & Pop & Unpop & Pop/Unpop\\
        \hline
        Appliances & 8.22 & 1.62 & 5.07 \\
        Baby Products & 39.63 & 4.76 & 8.32 \\
        Grocery and Gourmet Food & 31.55 & 4.59 & 6.87 \\
        Movies and TV & 24.33 & 4.50 & 5.41 \\
        Office Products & 21.25 & 3.97 & 5.35 \\
        Patio Lawn and Garden & 23.59 & 4.10 & 5.75 \\
        Pet Supplies & 66.22 & 5.93 & 11.17 \\
        Sports and Outdoors & 20.35 & 3.86 & 5.27 \\
        Toys and Games & 20.99 & 4.08 & 5.15 \\
      \toprule
    \end{tabular}
    }
    \label{tab:degree}
\end{table}

\subsection{Hyperparameters}
For the recommender system backbone, although hyperparameter tuning could potentially improve performance, we opted not to tune them due to the large number of experiments conducted. Instead, we fixed the hyperparameters across all experiments: the learning rate is set to $1e^{-4}$, batch size is set to 2048, embedding dimension is set to 64, l2 regularization weight is set to $1e^{-4}$. The number of hops for LightGCN is set to 2. Training will continue for up to 1000 epochs unless early stopping criteria are met, which occur when the best validation results do not change for 50 epochs. Note that the same set of parameters and training strategies are applied for the compared augmentation methods.

\subsection{Infrastructure}
All experiments are performed on a machine
with A100-80G GPU RAM and 128GB CPU RAM.

\end{document}